\documentclass[
    aps,
    prl,
    twocolumn,
    superscriptaddress,
    showpacs,
    floatfix
]{revtex4-2}

\usepackage{graphicx, amsmath, amssymb}
\usepackage{hyperref}
\usepackage{dsfont}
\usepackage{natbib}
\usepackage{kotex}

\usepackage{algorithm, algpseudocode}

\usepackage{xcolor}
\usepackage{tcolorbox}

\newcommand{\dd}{\textrm{d}}
\let\l\left
\let\r\right

\usepackage{empheq}
\usepackage{MnSymbol}

\begin{document}
\title{
Competing heterogeneities shape ordering via higher-order interactions
}

\author{Gangmin Son}
\email{gmson102@gmail.com}
\affiliation{School of Computational Sciences, Korea Institute for Advanced Study, Seoul 02455, Korea}

\author{Federico Battiston}
\email{battistonf@ceu.edu}
\affiliation{Department of Network and Data Science, Central European University, Vienna, Austria}
\affiliation{Department of AI, Data and Decision Sciences, Luiss University of Rome, Viale Romania 32, 00197, Rome, Italy}

\author{Deok-Sun Lee}
\email{deoksunlee@kias.re.kr}
\affiliation{School of Computational Sciences, Korea Institute for Advanced Study, Seoul 02455, Korea}
\affiliation{Center for AI and Natural Sciences, Korea Institute for Advanced Study, Seoul 02455, Korea}

\author{K.-I. Goh}
\email{kgoh@korea.ac.kr}
\affiliation{Department of Physics, Korea University, Seoul 02841, Korea}

\date{\today}
\begin{abstract}
Higher-order interactions admit richer structural heterogeneity than pairwise networks.
To understand how heterogeneity impacts collective phenomena we develop a framework based on the cavity method and apply it to the simplicial Ising model on heterogeneous hypergraphs. Unlike in homogeneous structures, group size and node degree play fundamentally different roles: size heterogeneity sharpens the transition via large-group unanimity, while degree heterogeneity softens it as hubs cooperatively seed ordering with non-hubs. 
Under either type of heterogeneity, continuous--discontinuous double transitions can arise, where the symmetry-breaking continuous transition is driven by pairs or by hubs, respectively.
When both heterogeneities coexist, cross-order degree correlations further modulate the phase diagram,
with anticorrelation delaying the group-driven discontinuous jump and broadening the hysteretic region.
Our results reveal the intricate interplay between size and degree heterogeneities in collective phenomena beyond pairwise interactions.
\end{abstract}
\maketitle


\textit{Introduction.---}Interactions in complex systems frequently occur not just between pairs of elements, but also among groups of three or more units. These higher-order interactions (HOIs)~\cite{battiston2020networks,bianconi2021higher,boccaletti2023structure} are essential for driving collective dynamics ranging from social adoption to neural synchrony. Crucially, HOIs are not reducible to superpositions of pairwise interactions; they introduce qualitatively new macroscopic phenomena, most prominently the emergence of abrupt phase transitions that are absent in the corresponding pairwise systems~\cite{iacopini2019simplicial,centola2007complex,centola2010spread}.

While the role of network topology, such as degree heterogeneity, in reshaping pairwise dynamics is well characterized~\cite{bianconi2002mean,dorogovtsev2002ising,leone2002ferromagnetic,dorogovtsev2008critical}, how the topology of HOI networks determines phase transitions remains largely an open question. Real-world hypergraphs are heterogeneous in both individual connectivity (node degree) and group size (hyperedge cardinality)~\cite{st2022influential}, and they can exhibit structural correlations with no pairwise analog. For instance, hubs naturally split into \textit{group-hubs} (nodes participating in many large groups) and \textit{pair-hubs} (nodes involved in many pairwise links), and the two roles need not coincide. Whether and how such properties---size heterogeneity, degree heterogeneity, and their cross-correlations---govern phase transitions in HOI networks has not been systematically addressed, largely because it requires going beyond standard mean-field treatments.

To close this gap, we develop a cavity-method framework for higher-order dynamics on random sparse hypergraphs with arbitrary heterogeneity, and apply it to the simplicial Ising model (SIM)~\cite{robiglio2025synergistic,son2024phase,robiglio2025higher} as an analytically tractable testbed.
The SIM captures group reinforcement through the simplicial rule, where an individual adopts a state only if all other members of its group share that state~\cite{iacopini2019simplicial}.
Its equilibrium formulation admits a systematic treatment on locally treelike hypergraphs, bypassing the analytical bottlenecks typical of nonequilibrium models.
This tractability enables a theoretically precise understanding of how heterogeneity alters collective phenomena in higher-order systems.

Here, we comprehensively map out the phase diagrams across arbitrary size and degree distributions, and find a rich phenomenology: size and degree heterogeneities---which enter symmetrically in the homogeneous limit---play opposite roles, with size heterogeneity sharpening the transition through large-group dominance and degree heterogeneity softening it through hub-induced nucleation. Both can produce continuous--discontinuous double transitions, but through distinct mechanisms.
Finally, we identify the correlation between group-level and pair-level degrees of the same node, the cross-order degree correlation (CODC)~\cite{zhang2023higher}, as an additional key structural variable that affects the transition scenario: negative CODC delays group-driven 
ordering and broadens the hysteretic region.

\begin{figure}[tb!]
    \centering
    \includegraphics[width=\linewidth]{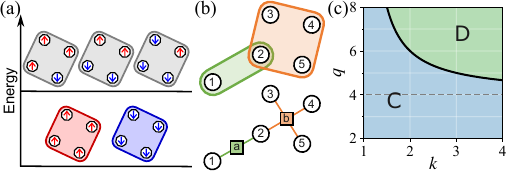}
    \caption{Model and role of node degree and hyperedge size. (a) Interaction rule in the simplicial Ising model (SIM). Possible spin configurations for a four-body group are shown along the energy levels. All non-unanimous states share the same excited-state energy, so that ordering requires the simultaneous alignment of all members. (b) A hypergraph (top) and its factor graph representation (bottom). Each hyperedge (shaded enclosure) of nodes (circle, top) can be mapped into a factor (filled square, bottom) connected to the nodes, i.e., $\partial a = \{1, 2\}$ and $\partial b = \{2,3,4,5\}$. (c) Phase diagram of the SIM on $k$-regular $q$-uniform hypergraphs. The solid line is the tricritical point (TP) line $(q-4)(k-1)=2$ separating continuous (C) and discontinuous (D) regimes.}
    \label{fig1}
\end{figure}

\textit{Formalism.---}In the SIM, Ising spins ${S}_i\in \{\pm 1\}$ reside on nodes $i \in \mathcal{V}=\{1,\ldots,N\}$ of a hypergraph where a hyperedge can involve more than two nodes~[Fig.~\ref{fig1}(a)].
Here, we adopt the factor graph representation, in which a hypergraph is mapped into a bipartite graph $(\mathcal{V},\mathcal{F},\mathcal{E}')$ of variable nodes (shortly, nodes) $i\in\mathcal{V}$, factor nodes (shortly, factors) $a\in\mathcal{F}$, and edges between them $(i,a)\in\mathcal{E}'$~[Fig.~\ref{fig1}(b)].
We denote the set of neighboring nodes (factors) of a factor (node) $a$ ($i$) by $\partial a$ ($\partial i$) respectively; then $\partial a$ represents a hyperedge.
With this notation, the Hamiltonian is written as
\begin{align}
    \label{eqn:hamiltonian}
    \mathcal{H}(\{{S}_i\})
    &= -\sum_{ a \in \mathcal{F}} J_{|\partial a|}
    \delta_{\{{S_i}\}_{i\in\partial a}}
    - H\sum_{i\in \mathcal{V}}{{S}_i},
\end{align}
where $J_{|\partial a|}>0$ is a ferromagnetic coupling that depends on the hyperedge size $|\partial a|$, $H$ is an external field, and the higher-order Kronecker delta enforces the simplicial rule:
\begin{align}
    \label{eqn:delta}
    \delta_{\{S_i\}_{i\in \partial a}}
    &= \begin{cases}
    1 & \text{if all ${S}_i$ are identical for $i\in\partial a$},\\
    0 & \text{otherwise},
    \end{cases}
\end{align}
as illustrated in Fig.~\ref{fig1}(a).
Since all non-unanimous configurations carry the same energy, the ordering constraint within a hyperedge becomes increasingly demanding as the group size grows.

On a locally treelike hypergraph, the cavity method~\cite{mezard2009information} yields a set of exact self-consistent equations for this equilibrium model, providing analytical access to the sparse regime only partially addressed in prior work~\cite{son2024phase,robiglio2025higher}.
An auxiliary variable $u_{a\to i}$, called the cavity bias, is introduced for each node--factor edge $(i,a)\in\mathcal{E}'$, representing the effective field that factor $a$ exerts on node $i$.
The self-consistent equations for $\{u_{a\to i}\}$ are given by
\begin{align}
    \label{eqn:sce}
    u_{a\to i}
    = \hat{u}(\{h_{j\to a}\}_{j\in\partial a\setminus i}),
\end{align}
where $h_{j\to a}=\sum_{b \in \partial j \setminus a} u_{b\to j}$
is the cavity field on node $j$ in the absence of factor $a$,
and $\hat{u}$ maps the cavity fields of the remaining $q-1$ neighbors to the bias exerted on node $i$, obtained by tracing out those spins (see End Matter).
The total magnetization $m\equiv\frac{1}{N}\sum_{i\in\mathcal{V}}
\langle S_i\rangle$ is then given by
\begin{align}
    \label{eqn:mag}
    m = \frac{1}{N}\sum_{i\in\mathcal{V}}\tanh{\left(
    \beta\sum_{b\in\partial i}u_{b\to i} + \beta H\right)}.
\end{align}

For $q$-uniform $k$-regular hypergraphs, since $u_{a\to i}=u$
for all node--factor pairs, Eq.~(\ref{eqn:sce}) reduces to
a single equation:
\begin{align}
    u &= \frac{1}{\beta}\tanh^{-1}{\frac{\mathcal{S}}
    {\mathcal{C}_1+\mathcal{C}_2}},
\end{align}
where
\begin{equation}
\begin{aligned}
    \mathcal{S} &=(e^{\beta J_q} - 1)
    \sinh{(\beta (q-1)(k-1)u)},\\
    \mathcal{C}_1 &= (e^{\beta J_q} - 1)
    \cosh{(\beta (q-1)(k-1)u)},\\
    \mathcal{C}_2 &= (2 \cosh{(\beta(k-1)u)})^{q-1}.
\end{aligned}
\end{equation}
Depending on $q$ and $k$, the transition can be either continuous
or discontinuous. In the na\"ive mean-field limit $k\to\infty$, the tricritical point (TP) occurs at $q=4$.
For finite $k$, the TP condition generalizes to
\begin{align}
    \label{eqn:tp}
    (q-4)(k-1) = 2,
\end{align}
so that $q$ and $k$ enter on equal footing in determining the TP line, as shown in Fig.~\ref{fig1}(c) [see Supplemental Material (SM), Sec.~S1~\cite{sm}].
This equivalence, however, breaks down once heterogeneity is
introduced, as we show in what follows.

Unlike pairwise networks, hypergraphs admit heterogeneity not only in node degree but also in group size (hyperedge cardinality or factor degree), and further allow the CODC under both degree and size heterogeneities.
For heterogeneous hypergraphs, the cavity bias $u_{a\to i}$ and field $h_{i\to a}$ become random variables whose distributions satisfy distributional self-consistency equations.
We consider two classes of hypergraphs:
(i)~random hypergraphs, constrained by arbitrary total degree and size distributions $P(k)$ and $Q(q)$; and
(ii)~random multiplex hypergraphs~\cite{bianconi2024theory},
constrained by the degree distribution separately for each hyperedge size, characterized by a distribution $P(\mathbf{k})$ of generalized-degree vectors $\mathbf{k} = (k_1, \ldots, k_q)$,
where $k_q$ denotes the number of size-$q$ hyperedges incident
to a node.

Near the critical point, the cavity bias distribution collapses to a delta function, rendering the self-consistent equations analytically tractable; this is the effective medium approximation (EMA)~\cite{dorogovtsev2008critical}, which we adopt throughout. Details are given in End Matter.

\begin{figure}[tb!]
    \centering
    \includegraphics[width=\linewidth]{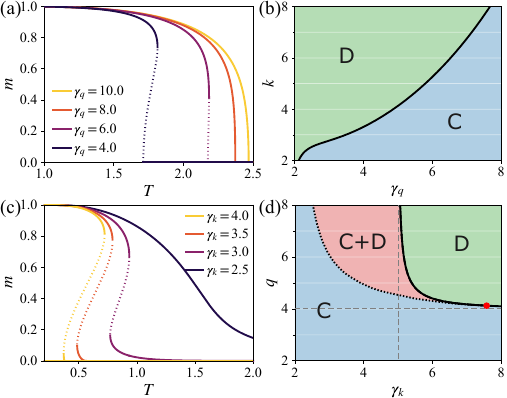}
    \caption{
    Phase transitions on heterogeneous hypergraphs.
    [(a),(c)] $m(T)$ for size- (a) and degree-heterogeneous (c) cases, with varying $\gamma_q$ (at $k=7$) and $\gamma_k$ (at $q=7$).
    [(b),(d)] Phase diagrams in the $(\gamma_q,k)$ (b) and $(\gamma_k,q)$ (d) planes.
    In (d), in addition to C and D regimes, a regime for continuous--discontinuous double transitions (C+D) appears.
    The black dotted line is the boundary between the C and (C+D) regions, the red circle marks the special tricritical point (STP) at which the (C+D) region emerges, and the gray dashed lines show the asymptotes of the TP line, $q=4$ and $\gamma_k=5$, respectively.
    }
    \label{fig2}
\end{figure}

\textit{Roles of size and degree heterogeneities.---}We first investigate size heterogeneity by considering $k$-regular hypergraphs with a power-law size distribution $Q(q)\propto q^{-\gamma_q}$ in $q_{\min}\le q \le q_{\max}$.
Figure~\ref{fig2}(a) shows the phase transitions for $k=7$ and various $\gamma_q$ ($q_{\min}=4$ and $q_{\max}=100$). As $\gamma_q$ decreases, i.e., the size heterogeneity increases, the transition `sharpens,' i.e., its type changes from continuous to discontinuous and its jump size and hysteresis width increase.
The phase diagram, shown in Fig.~\ref{fig2}(b), also illustrates that the continuous region becomes narrower as the size heterogeneity increases.
In addition, we investigate the cutoff dependence of the TP line and show that the minimum cutoff is more relevant than the maximum (see SM, Sec.~S2~\cite{sm}).

Conversely, degree heterogeneity plays an opposing role in phase transitions.
We consider $q$-uniform hypergraphs with a power-law degree distribution $P(k)\propto k^{-\gamma_k}$ in $k_{\min}\le k \le k_{\max}$, and exploit the annealed approximation, which is exact in $k_{\min}\to\infty$, for its technical advantage (see SM, Sec.~S3~\cite{sm}).
As shown in Fig.~\ref{fig2}(c), the transition `softens' as $\gamma_k$ decreases ($q=7$). Not only the type, jump size, and hysteresis, but also the critical exponents change. As in the conventional Ising model on scale-free networks~\cite{dorogovtsev2002ising}, the critical exponents depend on $\gamma_k$. For example, the magnetization exponent is $\beta_m = 1/(\gamma_k-3)$ for $3<\gamma_k<5$, and the critical temperature diverges for $\gamma_k \le 3$ (see SM, Sec.~S1~\cite{sm}). 

Furthermore, additional discontinuous transition can happen after a continuous transition, e.g., for $\gamma\in\{3.0, 3.5, 4.0\}$ in Fig.~\ref{fig2}(c).
Figure~\ref{fig2}(d) shows the continuous (C), discontinuous (D), and double transition (C+D) regions in the $(\gamma_k,q)$-plane.
The D region is bounded by a TP line with asymptotes $q=4$ as $\gamma_k\to\infty$ and $\gamma_k=5$ as $q\to\infty$.
The `special' TP (STP, red circle) indicates the emergence of (C+D) region (see SM, Sec.~S1~\cite{sm}).
Moreover, in contrast to the size-heterogeneous case, the cutoff dependence of the TP line suggests that the maximum cutoff is more relevant than the minimum (see SM, Sec.~S2~\cite{sm}).
To summarize, size and degree heterogeneities play complementary roles in shaping phase transitions, with the head and tail of the distributions being the relevant parts, respectively.

\textit{Size- and degree-specific magnetization in binary mixtures.---}To understand the mechanisms behind these heterogeneity effects, we analyze binary mixtures that isolate each type of heterogeneity in its minimal form.

\begin{figure}[tb!]
    \centering
    \includegraphics[width=\linewidth]{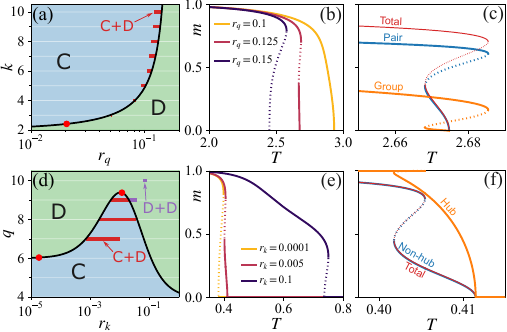}
    \caption{
    Phase transitions on $(2,10)$-biuniform $k$-regular [(a)--(c)] and $q$-uniform $(2,10)$-biregular [(d)--(f)] hypergraphs.
    [(a),(d)]~Phase diagrams in the $(r_q, k)$ (a) and $(r_k,q)$ (d) planes.
    Red horizontal bars represent (C+D) region.
    Purple horizontal bars represent a region for double discontinuous transitions (D+D).
    [(b),(e)]~$m(T)$ for several $r_q$ at $k=10$ (b) and several $r_k$ at $q=7$ (e).
    [(c),(f)]~Size-specific (c) and degree-specific (f) magnetizations. (c) $m_{\mathrm{pair}}$, $m_{\mathrm{group}}$, and total $m$ for $r_q=0.125$ from (b). (f) $m_{\mathrm{non\text{-}hub}}$, $m_{\mathrm{hub}}$, and total $m$ for $r_k=0.005$ from (e); $m$ is almost hidden behind $m_{\mathrm{non\text{-}hub}}$.
    }
    \label{fig3}
\end{figure}

We first consider $(2,q)$-biuniform $k$-regular hypergraphs, where interaction sizes can be either $2$ (pairs) or $q>2$ (groups).
The total degree distribution is $P(k')=\delta(k'-k)$ and the size distribution is $Q(q') = (1-r_q) \delta(q'-2) + r_q \delta(q'-q)$, where $r_q$ is the group fraction.
Figures~\ref{fig3}(a--c) show the results for $q=10$.
As $r_q$ increases, the transition changes from continuous to discontinuous, consistent with the results in Figs~\ref{fig2} (a--b).
Near the TP on the continuous side, the (C+D) double-transition regime appears [Fig.~\ref{fig3}(a)], bounded by the STP (red circle); this regime broadens with increasing $k$.
To isolate the role of each size of interactions, we define the size-specific magnetization $m_{\textrm{pair}}$ and $m_{\textrm{group}}$, the response each interaction type would produce alone at its current effective field strength
(see End Matter for more details).
As shown in Fig.~\ref{fig3}(c), the pair part $m_{\textrm{pair}}$ rises continuously at high temperature, while the group part $m_{\textrm{group}}$ remains negligible until lower temperature at which it jumps discontinuously, reflecting the unanimity constraint effect of large groups.

We then turn to $(2,k)$-biregular $q$-uniform hypergraphs, where node degrees can be either $2$ (non-hubs) or $k>2$ (hubs).
The degree distribution is $P(k') = (1-r_k)\delta(k'-2) + r_k\delta(k'-k)$ and the size distribution is $Q(q') = \delta(q'-q)$, where $r_k$ is the hub fraction.
Figures~\ref{fig3}(d--f) show the results for $k=10$.
The TP line exhibits an inverted V-shape [Fig.~\ref{fig3}(d)]. In the limits $r_k=0$ and $r_k=1$, the system reduces to a homogeneous hypergraph at degree $2$ or $k$, recovering the corresponding tricritical values $q=6$ and $q=38/9\approx4.22$, respectively, from Eq.~(\ref{eqn:tp}).
The hump-shaped continuous regime at intermediate $r_k$ confirms the synergistic origin of hub-induced softening observed in the power-law case. Neither hubs nor non-hubs alone can sustain a continuous transition for these values of $q$, yet their coexistence enables one.
Two STPs (red circles) bound the (C+D) double-transition regime from either side, as shown in Fig.~\ref{fig3}(d).
The degree-specific magnetization, $m_{\textrm{non-hub}}$ or $m_{\textrm{hub}}$, is defined as the local magnetization of nodes with specific degree, $2$ or $k$ (see End Matter for more details).
As illustrated in Fig.~\ref{fig3}(f), the hub magnetization $m_{\textrm{hub}}$ rises sharply upon cooling and approaches unity prior to the discontinuous jump, confirming that hubs initiate the continuous transition while contributing minimally to the jump itself. Conversely, the non-hub component $m_{\textrm{non-hub}}$, which dominates the total magnetization, undergoes a more pronounced discontinuous jump and continues to increase thereafter.
In addition to continuous--discontinuous double transitions, the biregular case also supports double discontinuous (D+D) transitions [purple horizontal lines in Fig.~\ref{fig3}(d)]; details are given in SM, Sec.~S4~\cite{sm}.

\textit{Correlations between pair- and group-hubs.---}While the previous sections treated size and degree heterogeneities independently, in real hypergraphs a node's pair-level and group-level degrees can be correlated.
This cross-order degree correlation (CODC)~\cite{zhang2023higher} is directly incorporated into our cavity framework, as it is a within-node property that leaves between-node degree correlations intact.
It constitutes a unique higher-order feature with no pairwise analog, though conceptually reminiscent of interlayer degree correlations in multiplex networks~\cite{min2014network}.

To probe its effect, we combine the two binary-mixture setups into a single biuniform and biregular hypergraph and use the multiplex ensemble.
Sizes are either $2$ or $q>2$, and the generalized degree vector is $\mathbf{k}=(k_2,k_q)$, where $k_2$ and $k_q$ are the size-$2$ and size-$q$ degrees, respectively.
The CODC is controlled by a parameter $\alpha\in[-1,1]$; the joint degree distribution is
\begin{align}
    \label{eqn:CODC}
    P(\mathbf{k}) = \frac{1+\alpha}{2}\,P_{+}(\mathbf{k}) + \frac{1-\alpha}{2}\,P_{-}(\mathbf{k}),
\end{align}
where $P_{+}(\mathbf{k})= \frac{1}{2} \delta(k_2-\kappa_{+})\delta(k_q-\kappa_{+}) + \frac{1}{2} \delta(k_2-\kappa_{-})\delta(k_q-\kappa_{-})$ and $P_{-}(\mathbf{k})= \frac{1}{2} \delta(k_2-\kappa_{+})\delta(k_q-\kappa_{-}) + \frac{1}{2} \delta(k_2-\kappa_{-})\delta(k_q-\kappa_{+})$, with $\kappa_{+}>\kappa_{-}$ denoting the hub and non-hub degrees.
Under $P_{+}$, pair-hubs are also group-hubs; under $P_{-}$, pair-hubs are group-non-hubs.
At $\alpha=+1$, the two hub identities coincide; at $\alpha=-1$, they are maximally disjoint; $\alpha=0$ is the uncorrelated baseline.

\begin{figure}[tb!]
    \centering
    \includegraphics[width=\linewidth]{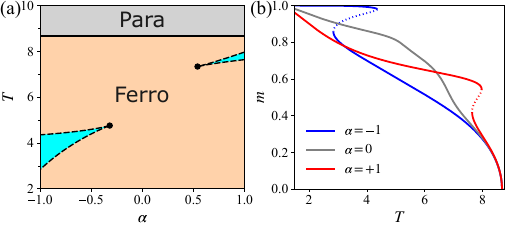}
    \caption{
    Effect of cross-order degree correlations (CODC) on $(2,20)$-biuniform $(2,20)$-biregular hypergraphs. (a) Phase diagram in the $(\alpha, T)$ plane, where $\alpha\in[-1,1]$ controls the CODC ($\alpha=+1$: pair-hubs and group-hubs coincide; $\alpha=-1$: pair-hubs and group-hubs are disjoint). The gray and orange regions denote the paramagnetic (Para) and ferromagnetic (Ferro) phases, respectively. The solid line marks the continuous transition between the two phases. Within the Ferro phase, the cyan region bounded by dashed lines indicates where an additional discontinuous jump and hysteresis occur. Black circles mark critical points where continuous transition occurs. (b) $m(T)$ for $\alpha=-1$, $0$, and $+1$.
    }
    \label{fig4}
\end{figure}

Figure~\ref{fig4}(a) shows the phase diagram in the $(\alpha, T)$ plane for $q=20$ and $\kappa_{+}=20$, $\kappa_{-}=2$. For $|\alpha|$ close to unity, the system exhibits a continuous--discontinuous double transition upon cooling, with hysteresis in the cyan region.
As $\alpha$ decreases from $+1$ to $-1$, this hysteretic region shifts to lower temperatures.
At intermediate $\alpha$, while no double transition occurs, the magnetization grows with nontrivially changing concavity upon cooling [Fig.~\ref{fig4}(b)].

The mechanism is as follows.
When pair-hubs and group-hubs coincide ($\alpha \approx 1$), the same nodes channel both pairwise and group-mediated ordering: Their large size-$2$ degree nucleates local order early, and their large size-$q$ degree amplifies it through the simplicial coupling.
The two mechanisms reinforce each other, so that the system rapidly enters the hysteretic regime.
When the two hub populations are disjoint ($\alpha \approx -1$), pairwise and group-mediated ordering decouple: Pair-hubs nucleate pairwise order at high temperatures but participate in few groups, while group-hubs lack the pairwise reinforcement needed to trigger group ordering.
Consequently, a much larger temperature gap separates the continuous onset from the discontinuous jump.
At intermediate $\alpha \approx 0$, all four node types (pair-hub/group-hub, pair-hub/group-non-hub, pair-non-hub/group-hub, pair-non-hub/group-non-hub) coexist in comparable proportions.
The absence of a dominant ordering channel leads to a gradual, monotonic growth of magnetization without a discontinuous jump.
These results show that the CODC, despite being the most elementary correlation within a node's generalized-degree vector, produces a rich variety of phase-transition scenarios.
This suggests that the CODC is a fundamental but so far overlooked structural variable in the theory of collective phenomena on higher-order structures.

\textit{Conclusions.---}We have shown that the simplicial Ising model on heterogeneous hypergraphs exhibits rich phase-transition phenomenology: size and degree heterogeneities play opposite roles, and cross-order degree correlations markedly reshape the resulting phase diagram featuring multiple transitions.
Our analytical framework provides a systematic route to characterize collective phenomena on hypergraphs with arbitrary heterogeneity and is not restricted to the SIM.
Analogous phenomenology may thus emerge in other higher-order interaction systems, including contagion and synchronization dynamics, and our work establishes a foundation for understanding heterogeneity effects across higher-order interactions more broadly.

\textit{Acknowledgements.---}This work was supported by KIAS Individual Grant [No. CG098002 (G.S.) and No. CG079902 (D.-S.L.)] at Korea Institute for Advanced Study; by Austrian Science Fund (FWF) through projects 10.55776/PAT1052824 and 10.55776/PAT1652425 (F.B.); and by National Research Foundation of Korea (NRF) grant funded by the Korea government (MSIT) [No. RS-2025-00558837] (K.-I.G.).

\bibliography{refs}


\onecolumngrid
\section*{End Matter}

\subsection{Cavity Equations}
\label{sec:sm_cavity}

\paragraph{General equations.} We derive the cavity equations for the simplicial Ising model (SIM) on a general locally treelike hypergraph.
In the factor graph representation, the hypergraph $(\mathcal{V},\mathcal{E})$
is mapped to a bipartite graph $(\mathcal{V},\mathcal{F},\mathcal{E}')$, where
each hyperedge $e\in\mathcal{E}$ of size $|e|=q$ becomes a factor $a\in\mathcal{F}$ of degree $q$.

For a factor $a$ of size $q=|\partial a|$ with coupling $J_{q}$,
the cavity bias $u_{a\to i}$ is defined by
\begin{align}
    e^{2\beta u_{a\to i}} \equiv
    \frac{Z_a(S_i=+1)}{Z_a(S_i=-1)},
\end{align}
where
\begin{align}
    Z_a(S_i) = \sum_{\{S_j\}_{j\in\partial a\setminus i}}
    \exp\left[\beta J_{q}\,\delta_{\{S_l\}_{l\in\partial a}}
    + \beta\sum_{j\in\partial a\setminus i} h_{j\to a}\, S_j\right]
\end{align}
is the partial partition function by summing over all spins in a hyperedge $\partial a$ except $S_i$, and $h_{j\to a} = \sum_{b\in\partial j\setminus a} u_{b\to j}$ is the cavity field on node $j$ in the absence of factor $a$.

For $S_i=+1$, the $2^{q-1}$ spin configurations of the remaining $q-1$ nodes split into two classes:
(i)~the unanimous state $S_j=+1$ for all $j$ (one configuration), which carries the simplicial energy $-J_{q}$ and contributes
$e^{\beta J_{q}} \prod_{j\in\partial a\setminus i} e^{\beta h_{j\to a}}$;
(ii)~all other $2^{q-1}-1$ non-unanimous configurations, which carry zero simplicial energy and contribute $\prod_{j} 2\cosh(\beta h_{j\to a}) - \prod_{j} e^{\beta h_{j\to a}}$.
Combining, we obtain
\begin{align}
    Z_a(S_i=+1) = J'_{q}\,e^{\beta\sum_j h_{j\to a}}
    + \prod_{j\in\partial a\setminus i} 2\cosh(\beta h_{j\to a}),
\end{align}
where $J'_{q}\equiv e^{\beta J_{q}}-1$.
By $\mathbb{Z}_2$ symmetry ($S_i=-1$ makes the all-minus state unanimous),
$Z_a(S_i=-1) = J'_{q}\,e^{-\beta\sum_j h_{j\to a}}
+ \prod_{j} 2\cosh(\beta h_{j\to a})$.
Therefore, the function $\hat{u}$ in the main text Eq.~(3) takes the form
\begin{align}
    \label{eqn:sm_uhat}
    \beta\,\hat{u}(\{h_{j\to a}\}_{j\in\partial a\setminus i})
    = \tanh^{-1}\left[
    \frac{J'_{q}\sinh\bigl(\beta\sum_j h_{j\to a}\bigr)}
    {J'_{q}\cosh\bigl(\beta\sum_j h_{j\to a}\bigr)
    + \prod_j 2\cosh(\beta h_{j\to a})}
    \right],
\end{align}
where all sums and products run over $j\in\partial a\setminus i$.
At $h_{j\to a}=0$ for all $j$, the numerator vanishes and $\hat{u}=0$; the
disordered state is always a fixed point.
Setting $q=2$ recovers the standard Ising cavity equation
$\beta\,\hat{u}=\tanh^{-1}[\tanh(\beta J_2/2)\,\tanh(\beta h)]$.

\paragraph{Random hypergraphs.}
For a random hypergraph with degree distribution $P(k)$ and size
distribution $Q(q)$, subject to
$N\langle k\rangle = |\mathcal{E}|\langle q\rangle$, the cavity biases
become i.i.d.\ random variables with distributions
$\mathcal{Q}(u)$ and $\mathcal{P}(h)$ satisfying
\begin{align}
    \label{eqn:sm_dsce_Q}
    \mathcal{Q}(u) &= \sum_{q} \tilde{Q}(q)
    \int \prod_{j=1}^{q-1} \dd h_j\,\mathcal{P}(h_j)\;
    \delta\bigl(u - \hat{u}(\{h_j\})\bigr), \\
    \label{eqn:sm_dsce_P}
    \mathcal{P}(h) &= \sum_{k} \tilde{P}(k)
    \int \prod_{b=1}^{k-1} \dd u_b\,\mathcal{Q}(u_b)\;
    \delta\Bigl(h - \textstyle\sum_{b=1}^{k-1} u_b\Bigr),
\end{align}
where $\tilde{Q}(q)\equiv qQ(q)/\langle q\rangle$ and $\tilde{P}(k)\equiv kP(k)/\langle k\rangle$ are the probabilities that a randomly chosen edge leads to a factor of size $q$ and a node of degree $k$, respectively.
The magnetization is
\begin{align}
    m = \sum_k P(k) \int \prod_{b=1}^{k}\dd u_b\,\mathcal{Q}(u_b)\;
    \tanh\Bigl(\beta\textstyle\sum_{b=1}^{k} u_b + \beta H\Bigr).
\end{align}

Setting $\mathcal{Q}(u)=\delta(u-\bar{u})$ collapses
Eq.~(\ref{eqn:sm_dsce_P}) to
$\mathcal{P}(h)=\sum_k \tilde{P}(k)\,\delta(h-(k-1)\bar{u})$.
The self-consistent equation becomes
\begin{align}
    \label{eqn:sm_ema_rh}
    \bar{u}
    &=\sum_{q}\tilde{Q}(q)\sum_{k_1}{\tilde{P}(k_1)} \cdots \sum_{k_{q-1}}{\tilde{P}(k_{q-1})} \hat{u}\bigl(\{(k_j-1)\bar{u}\}\bigr)\\
    &\equiv
    \left\llangle
    \hat{u}\bigl(\{(k_j-1)\bar{u}\}\bigr)
    \right\rrangle,
\end{align}
where $\llangle\cdot\rrangle$ denotes the average over $\tilde{Q}(q)$ and $(q-1)$ independent $\tilde{P}(k)$.
The magnetization simplifies to
$m = \sum_k P(k)\,\tanh(\beta k \bar{u} + \beta H)$.
This effective medium approximation (EMA) is exact near the critical point, where $\bar{u}\to 0$ and the
distribution $\mathcal{Q}(u)$ collapses to $\delta(u)$. At lower temperatures, the EMA agrees better with exact results when hypergraphs are more dense and homogeneous.

\paragraph{Random multiplex hypergraphs.} For the multiplex ensemble, the generalized-degree vector of node $i$ is
$\mathbf{k}_i=(k_i^{(q_1)},k_i^{(q_2)},\ldots)$, where $k_i^{(q)}$
denotes the number of size-$q$ hyperedges incident to $i$.
The generalized-degree distribution $P(\mathbf{k})$ determines the size
distribution via $N\langle k^{(q)}\rangle = q|\mathcal{E}_q|$, giving
$Q(q)=(\langle q\rangle/q)\,\langle k^{(q)}\rangle/\langle k\rangle$.
The cavity biases are stratified by interaction size:
$u^{(q)}$ and $h^{(q)}$ for each size $q$, with distributions
$\mathcal{Q}_q(u^{(q)})$ and $\mathcal{P}_q(h^{(q)})$ satisfying
\begin{align}
    \label{eqn:sm_dsce_mpx_Q}
    \mathcal{Q}_q(u^{(q)}) &=
    \int \prod_{i=1}^{q-1}\dd\mathcal{P}_q(h_i^{(q)})\;
    \delta\bigl(u^{(q)} - \hat{u}(\{h_i^{(q)}\})\bigr),\\
    \label{eqn:sm_dsce_mpx_P}
    \mathcal{P}_q(h^{(q)}) &=
    \sum_{\mathbf{k}} \frac{k^{(q)}\,P(\mathbf{k})}{\langle k^{(q)}\rangle}
    \int \prod_{q'}\left[
    \prod_{\alpha=1}^{k^{(q')}-\delta_{q,q'}}
    \dd\mathcal{Q}_{q'}(u_\alpha^{(q')})\right]
    \delta\Bigl(h^{(q)}
    - \sum_{q'}\sum_{\alpha} u_\alpha^{(q')}\Bigr),
\end{align}
where the biased generalized-degree distribution
$k^{(q)}P(\mathbf{k})/\langle k^{(q)}\rangle$ samples a neighbor reached
through a randomly chosen size-$q$ factor, and $k^{(q')}-\delta_{q,q'}$ accounts for the
removal of the factor through which the neighbor is reached.
The cavity field $h^{(q)}$ thus receives contributions from all layers,
not only layer $q$.
The magnetization is
\begin{align}
    m = \sum_{\mathbf{k}} P(\mathbf{k})
    \int \prod_{q}\left[\prod_{\alpha=1}^{k^{(q)}}
    \dd\mathcal{Q}_q(u_\alpha^{(q)})\right]
    \,\tanh\Bigl(\beta\sum_{q}\sum_{\alpha=1}^{k^{(q)}}
    u_\alpha^{(q)}\Bigr).
\end{align}

Setting $\mathcal{Q}_q(u^{(q)})=\delta(u^{(q)}-\bar{u}^{(q)})$
collapses Eq.~(\ref{eqn:sm_dsce_mpx_P}) to
\begin{align}
    \mathcal{P}_q(h^{(q)}) = \sum_{\mathbf{k}}
    \frac{k^{(q)}\,P(\mathbf{k})}{\langle k^{(q)}\rangle}\;
    \delta\Bigl(h^{(q)}
    - \sum_{q'} k^{(q')}\bar{u}^{(q')} + \bar{u}^{(q)}\Bigr).
\end{align}
Substituting into Eq.~(\ref{eqn:sm_dsce_mpx_Q}), the self-consistent
equations become
\begin{align}
    \label{eqn:sm_ema_mpx}
    \bar{u}^{(q)} =
    \left\llangle
    \,\hat{u}\bigl(\{\bar{h}_i^{(q)}\}_{i=1}^{q-1}\bigr)
    \right\rrangle_q,
    \qquad
    \bar{h}_i^{(q)} = \sum_{q'} k_i^{(q')}\bar{u}^{(q')} - \bar{u}^{(q)},
\end{align}
where $\llangle\cdot\rrangle_q$ averages over $q-1$ i.i.d.\ generalized
degrees $\mathbf{k}_i$ drawn from
$k^{(q)}P(\mathbf{k})/\langle k^{(q)}\rangle$.
The magnetization simplifies to
$m = \sum_{\mathbf{k}} P(\mathbf{k})\,
\tanh(\beta\sum_{q'} k^{(q')}\bar{u}^{(q')})$.
For the $(2,q)$-biuniform $(2,k)$-biregular case,
$\mathbf{k}=(k^{(2)},k^{(q)})$ and Eq.~(\ref{eqn:sm_ema_mpx}) reduces to two
coupled equations for $\bar{u}^{(2)}$ and $\bar{u}^{(q)}$.

\subsection{Magnetization Decomposition}
\label{sec:sm_decomp}

We discuss how to quantify the isolated roles of different node degrees or interaction sizes in the magnetization.

\paragraph{Size-specific magnetization.}
For $(2,q)$-biuniform $k$-regular hypergraphs, the size distribution is
$Q(q')=(1-r_q)\delta(q'-2)+r_q\delta(q'-q)$ and every node has the same
total degree $k$.
Under the EMA, the self-consistent equation has a single order parameter
$\bar{u}$, and the effective fields from a single pairwise and group
factor are, respectively,
\begin{align}
    \bar{u}_{\textrm{pair}} &\equiv
    \hat{u}\bigl((k-1)\bar{u}\bigr),\\
    \bar{u}_{\textrm{group}} &\equiv
    \hat{u}\bigl(
    \underbrace{(k-1)\bar{u},\ldots,(k-1)\bar{u}}_{(q-1)\textrm{\ terms}}\bigr).
\end{align}
A node of total degree $k$ participates in $\langle k^{(2)}\rangle = 2(1-r_q)k/\langle q\rangle$ pairwise and
$\langle k^{(q)}\rangle = qr_q k/\langle q\rangle$ group interactions on average, so the cavity bias decomposes exactly by interaction size:
\begin{align}
    k\bar{u} = \langle k^{(2)}\rangle\bar{u}_{\textrm{pair}}
             + \langle k^{(q)}\rangle\bar{u}_{\textrm{group}}.
\end{align}
We define the size-specific magnetizations
\begin{align}
    m_{\textrm{pair}} &\equiv
    \tanh(\beta \langle k^{(2)}\rangle \bar{u}_{\textrm{pair}}), \\
    m_{\textrm{group}} &\equiv
    \tanh(\beta \langle k^{(q)}\rangle \bar{u}_{\textrm{group}}),
\end{align}
representing the magnetization a node would sustain under each interaction type alone at its current effective field strength. Note that $m = \tanh(\beta k \bar{u}) \neq m_{\textrm{pair}} + m_{\textrm{group}}$
in general; the discrepancy reflects the nonlinear cooperativity between the two ordering channels.

\paragraph{Degree-specific magnetization.}
For $(2,k)$-biregular $q$-uniform hypergraphs with
$P(k')=(1-r_k)\delta(k'-2)+r_k\delta(k'-k)$, nodes partition into two degree classes, so the total magnetization decomposes exactly as
\begin{align}
    m = (1-r_k)\,m_2 + r_k\,m_k,
\end{align}
where $m_{k'} = \tanh(\beta k' \bar{u})$ is the per-node magnetization of degree-$k'$ nodes.
Near the continuous onset ($\bar{u}\to 0^+$),
${m_k}/{m_2} \to k/2$,
confirming that each hub is $k/2$ times more magnetized than each non-hub.

\clearpage
\onecolumngrid

\begin{center}
    \vspace*{0.5em}
    {\large\bfseries Supplemental Material}
\end{center}
\vspace{1.5em}

\setcounter{page}{1}
\setcounter{equation}{0}
\setcounter{figure}{0}
\setcounter{section}{0}
\renewcommand{\theequation}{S\arabic{equation}}
\renewcommand{\thefigure}{S\arabic{figure}}
\renewcommand{\thesection}{S\arabic{section}}
\setcounter{secnumdepth}{3}

\section{Phase Diagram Analysis}
\label{sec:sm_phase}

We expand $\hat{u}$ [Eq.~(\ref{eqn:sm_uhat}) of End Matter] as a power series in $h_j=k'_j\bar{u}$, with $q-1$ neighbors whose excess degrees are $k'_j\equiv k_j -1$. Only odd powers appear due to $\mathbb{Z}_2$ symmetry. Writing $J'_q\equiv e^{\beta J_q}-1$ and $S_n\equiv\sum_{j=1}^{q-1}(k'_j)^n$, the expansion $\beta\,\hat{u}=\sum_{n=1,3,5,\ldots}c_n\,\bar{u}^n$ has coefficients
\begin{align}
    c_1 &= \frac{J'_q}{J'_q+2^{q-1}}\,S_1, \label{eqn:sm_c1}\\[4pt]
    c_3 &= \frac{2^{q-1}J'_q}{6(J'_q+2^{q-1})^3}\,
    \bigl[(2^{q-1}-J'_q)S_1^3 - 3(J'_q+2^{q-1})S_1 S_2\bigr],
    \label{eqn:sm_c3}\\[4pt]
    c_5 &= \frac{2^{q-1}J'_q}{120(J'_q+2^{q-1})^5}\,\biggl[
    (2^{q-1}-J'_q)(2^{2(q-1)}-10\cdot 2^{q-1}J'_q+J'^2_q)\,S_1^5 \notag\\
    &\quad - 10(2^{3(q-1)}-3\cdot 2^{2(q-1)}J'_q-3\cdot 2^{q-1}J'^2_q+J'^3_q)\,S_1^3 S_2
    \notag\\
    &\quad + 5(J'_q+2^{q-1})^2\bigl(
    3(2^{q-1}-J'_q)S_1 S_2^2 + 2(J'_q+2^{q-1})S_1 S_4
    \bigr)\biggr]. \label{eqn:sm_c5}
\end{align}
These coefficients are obtained by expanding
\begin{align}
\beta\,\hat{u}=\tanh^{-1}{\l[\frac{J'_q\sinh(\sum k'_j\bar{u})}{J'_q\cosh(\sum k'_j\bar{u})+\prod_j 2\cosh(k'_j\bar{u})}\r]}
\end{align}
as a power series in $\bar{u}$ and collecting terms order by order.

Under the effectiev medium approximation (EMA), the self-consistent equation $\bar{u}=\llangle\,\hat{u}(\{k'_j\bar{u}\})\rrangle$ yields the effective Landau expansion
\begin{align}
    \bar{u} = \llangle c_1\rrangle\,\bar{u}
    + \llangle c_3\rrangle\,\bar{u}^3
    + \llangle c_5\rrangle\,\bar{u}^5 + \cdots,
\end{align}
where $\llangle\cdot\rrangle$ averages over the excess size distribution and the $(q-1)$ excess degree distributions. The continuous transition occurs at $\llangle c_1\rrangle=1$, determining $T^*$. The tricritical point (TP) is located by
\begin{align}
    \llangle c_1\rrangle = 1, \qquad \llangle c_3\rrangle = 0.
\end{align}
In the $q$-uniform $k$-regular case, these reduce to $(q-4)(k-1)=2$ [Eq.~(\ref{eqn:tp}) of the main text]. For heterogeneous hypergraphs, the tricritical point (TP) conditions are solved numerically for the phase diagrams in Fig.~\ref{fig2}.

The special tricritical point (STP) is located by $\llangle c_1\rrangle = 1$, $\llangle c_3\rrangle = 0$, $\llangle c_5\rrangle = 0$. At the STP, the leading nonlinearity is seventh order, signaling the onset of the (C+D) regime: a continuous branch $\bar{u}\propto(T^*-T)^{1/2}$ emerges at $T^*$, and at a lower temperature a saddle-node bifurcation produces a discontinuous jump. For the annealed approximation, the STP condition can be solved for continuous $q$, yielding the (C+D) regime boundary in Fig.~\ref{fig2}(d). For the binary mixtures [Figs.~\ref{fig3}(a) and~\ref{fig3}(d)], the STP is identified at each integer $k$ or $q$ by interpolating the sign change of $\llangle c_5\rrangle$ along the TP line.

For scale-free degree distributions $P(k)\propto k^{-\gamma_k}$ in $k\in[k_{\min},\infty)$ with $\gamma_k>3$, the critical temperature $T^*$ is finite and the critical exponent $\beta_m$ for $m\sim(T^*-T)^{\beta_m}$ is
\begin{align}
    \beta_m =
    \begin{cases}
        1/2 & \gamma_k > 5,\\
        1/(\gamma_k-3) & 3 < \gamma_k < 5.
    \end{cases}
\end{align}
For $\gamma_k\le 3$, $\langle k^2\rangle$ diverges and $T^*\to\infty$. These results parallel those of the standard Ising model on scale-free networks~\cite{bianconi2002mean,dorogovtsev2002ising,leone2002ferromagnetic}.

\section{Cutoff Dependence of the Tricritical-Point Line}
\label{sec:sm_cutoff}

\begin{figure}[tb!]
    \centering
    \includegraphics[width=0.5\linewidth]{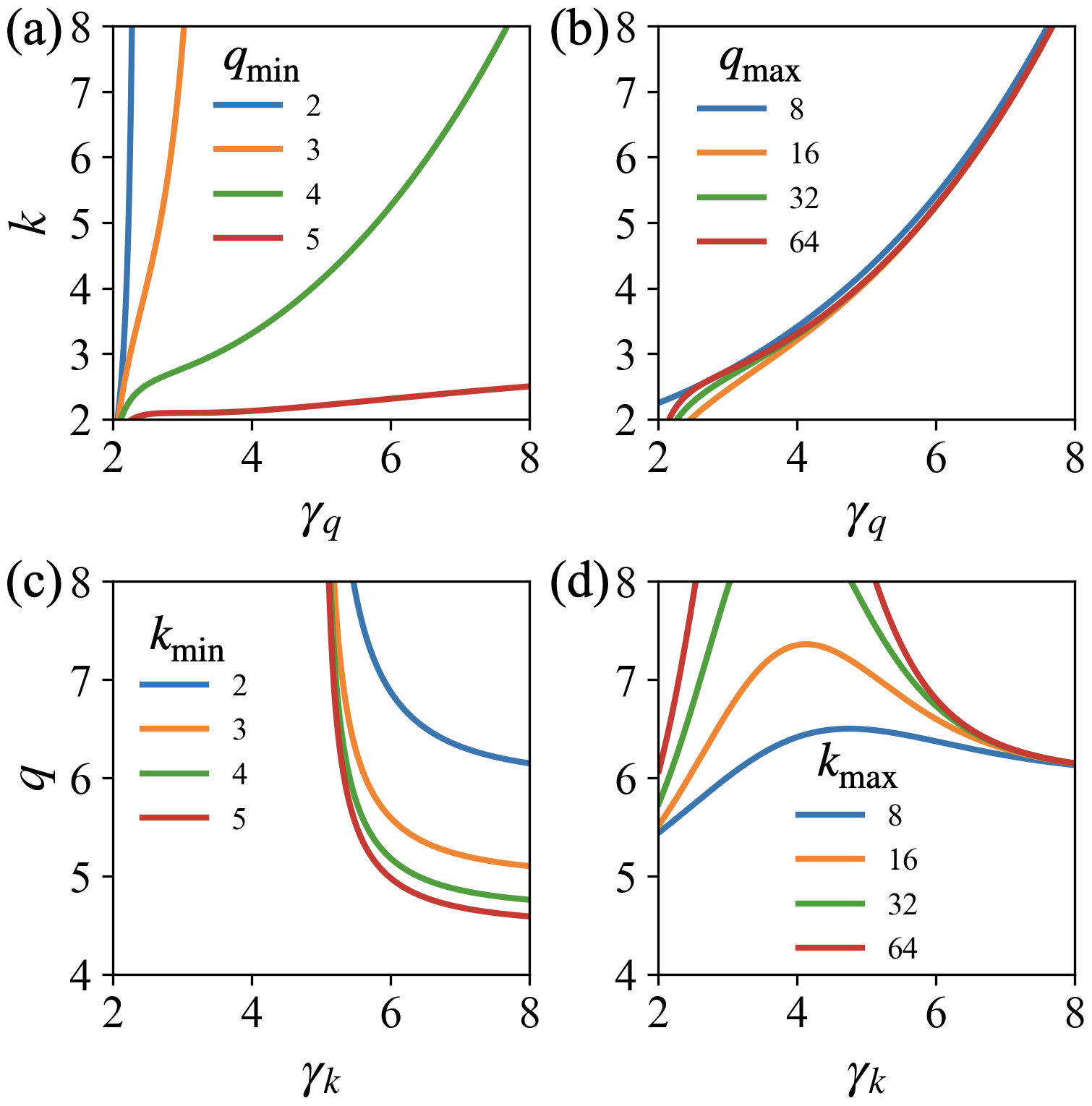}
    \caption{Tricritical-point (TP) lines for $k$-regular hypergraphs with power-law size distribution $Q(q)\sim q^{-\gamma_q}$ [(c),(d)] and for $q$-uniform hypergraphs with power-law degree distribution $P(k)\sim k^{-\gamma_k}$ [(a),(b)], varying the minimum [(a),(c)] and maximum [(b),(d)] cutoffs.
    The maximum cutoffs used in (a) and (c) are $q_{\max}=100$ and $k_{\max}\to\infty$, respectively. The minimum cutoffs used in (b) and (d) are $q_{\min}=4$ and $k_{\min}=2$, respectively.
    }
    \label{fig:cutoff}
\end{figure}

To assess how the cutoffs shape the phase diagram, we trace the TP line while varying each cutoff in turn.
For size heterogeneity, the TP line shifts appreciably with $q_{\min}$ but is nearly insensitive to $q_{\max}$ [Figs.~\ref{fig:cutoff}(a),(b)], showing that the `head' of $Q(q)$, rather than its tail, is relevant in shaping the phase transitions.
For degree heterogeneity, the situation reverses [Figs.~\ref{fig:cutoff}(c),(d)]: the TP line is far more sensitive to $k_{\max}$ than to $k_{\min}$. Finite $k_{\max}$ produces inverted-U-shaped TP lines whose low-$\gamma_k$ branch vanishes as $k_{\max}\to\infty$, consistent with Fig.~\ref{fig2}(d) of the main text.

\section{Annealed Approximation}
\label{sec:sm_annealed}

For $q$-uniform hypergraphs with degree distribution $P(k)$, we employ the annealed approximation~\cite{bianconi2002mean}, in which each node $i$ carries a fitness $k_i$ and the probability that nodes $\{i_1,\ldots,i_q\}$ share a hyperedge is proportional to $\prod_{l=1}^{q}k_{i_l}$.
The effective Hamiltonian for $q$-uniform hypergraphs is
\begin{align}
    \mathcal{H}(\{S_i\})
    = -J_q\sum_{e\in\mathcal{E}} w_e\,\delta_{\{S_i\}_{i\in e}},
    \qquad
    w_e = \frac{k_{i_1}\cdots k_{i_q}}{(N\langle k\rangle)^{q-1}},
\end{align}
where $\mathcal{E}$ is the set of hyperedges,
and the degree-weighted magnetization is
$M = \sum_i k_i\langle S_i\rangle / (N\langle k\rangle)$.
A mean-field decoupling yields the self-consistent equation
\begin{align}
    \label{eqn:sm_annealed_sce}
    M = \sum_k \frac{k}{\langle k\rangle}\,P(k)m_k
\end{align}
where
\begin{align}
    m_k = \tanh\left(\frac{1}{2}\beta k J_q\left[
    \left(\frac{1+M}{2}\right)^{q-1}
    - \left(\frac{1-M}{2}\right)^{q-1}
    \right]\right).
\end{align}
This equation can also be obtained as the $k_{\min}\to\infty$ limit of the EMA on the random hypergraph ensemble [Eq.~(\ref{eqn:sm_ema_rh})].
We can calculate the unweighted magnetization $m$ from $M$ by
\begin{align}
    m = \sum_{k}P(k) m_k.
\end{align}
In practice, we replace the sum over $k$ by an integral for continuous degree distributions.

\section{Double Discontinuous (D+D) Transition in Biregular Hypergraphs}
\label{sec:sm_biregular}

\begin{figure}[tb!]
    \centering
    \includegraphics[width=0.5\linewidth]{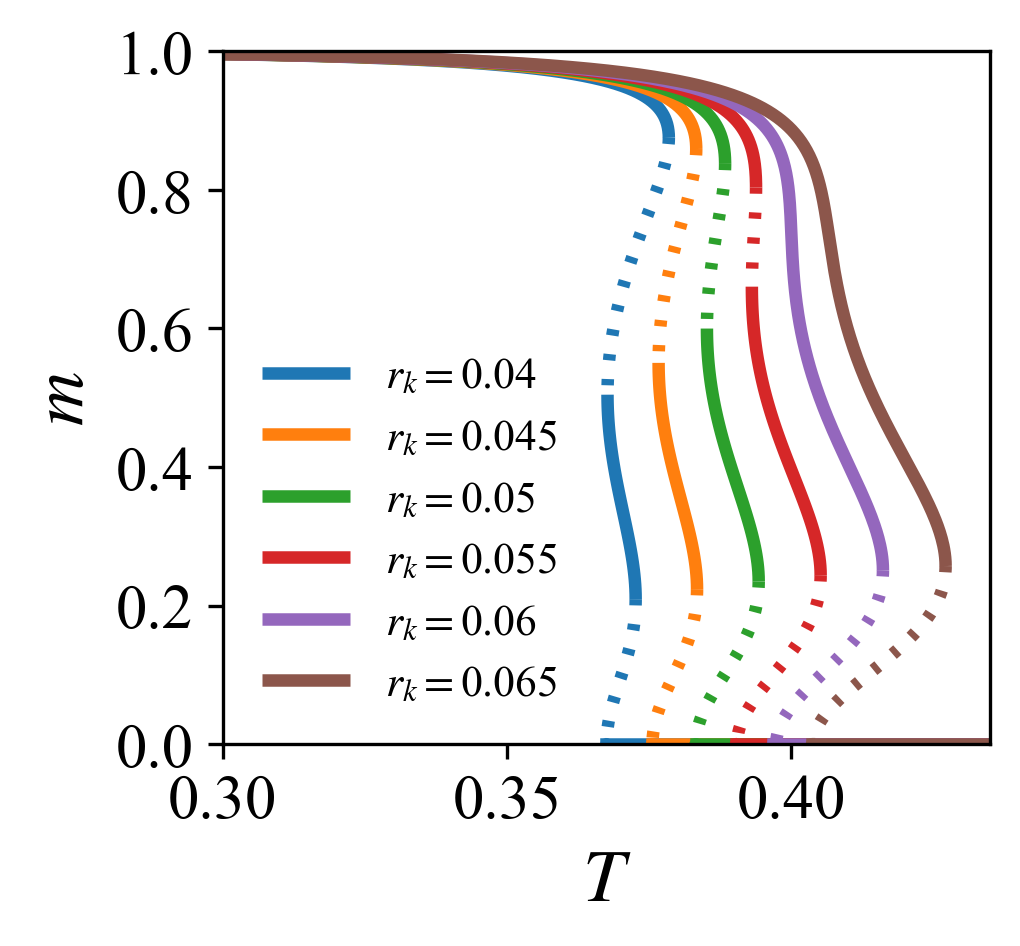}
    \caption{$m(T)$ in the (D+D) regime in Fig.~3(d) for $q=9$. The cases for $r_k=0.06$ and $0.065$ are outside of this regime.}
    \label{fig:nontrivial_regime}
\end{figure}

In the $(2,k)$-biregular $q$-uniform case with $P(k')=(1-r_k)\delta(k'-2)+r_k\delta(k'-k)$, in the ordinary discontinuous regime [green region in Fig.~\ref{fig3}(d)], the $m(T)$ curve has one turning point for $m>0$, where the stable branch (solid) meets the unstable branch (dashed), producing a single discontinuous jump. In the (D+D) region, marked by narrow purple horizontal bars in Fig.~\ref{fig3}(d), three such turning points appear for $m>0$ (Fig.~\ref{fig:nontrivial_regime}), enabling double discontinuous jumps.

\end{document}